\documentclass[conference]{IEEEtran}

\usepackage{graphicx}
\usepackage{amsmath,amssymb}
\usepackage{newtxtext,newtxmath}
\usepackage{booktabs}
\usepackage{array}
\usepackage{xspace}
\usepackage{xurl}
\usepackage{placeins}
\usepackage{capt-of}
\usepackage{balance}

\graphicspath{{figures/}}

\newcommand{\aid}{AID-Guard\xspace}

\begin{document}

\title{AID-Guard: Stateful Authorization for Delegated Agent Effects}

\author{
\IEEEauthorblockN{Yingzhe Tong, Leyu Dai, and Songhui Guo}
\IEEEauthorblockA{Information Engineering University\\
Emails: \{tyingzhe@outlook.com, tsubasa512@163.com, songhui.guo@outlook.com\}}
}

\maketitle

\begin{abstract}
Tool-using AI agents turn delegated tasks into provider effects, yet
authorization commonly ends at admission while mutable provider state,
delivery, retry, and recovery continue. A request may change before commit, or
an effect may occur while its response is lost; a replacement can then create a
second effect from one approval. We present AID-Guard, a stateful
authorization-to-effect closure protocol. It revalidates the exact approved
request and current provider state at commit, retains one reservation while the
outcome is ambiguous, and permits release or one successor only after a
terminal result or certified no effect with a delivery fence. For supported
provider contracts, one reservation therefore yields at most one effect across
retry and recovery. To our knowledge, AID-Guard is the first evaluated
agent-authorization protocol to preserve one reservation lineage across
provider commit and ambiguity, permitting at most one successor only after
certified no effect and provider-terminal predecessor fencing through the
declared recovery horizon.

We implement AID-Guard as a Python/SQLite prototype and evaluate it in a
declared loopback MCP domain. Thirteen live mutations produced no unauthorized
provider effects, four outcome traces preserved the specified commit and
recovery semantics, three concurrent histories were linearizable, and complete
evidence bundles supported public verification and privileged replay. All 210
Stripe provider-contract trials matched their predeclared outcomes. Across
two SaaS providers, 40 terminalize-then-successor schedules, 30 overlapping
Stripe confirm/cancel races, and 10 Stripe crash-recovery schedules completed
without a duplicate effect. Under complete proposer compromise, AID-Guard blocked all 44
submitted attacks while admitting all 44 matched legitimate proposals. The
strict exact-manifest profile reduced benign utility by 35.4--43.8 percentage
points; in a separate no-seed frontier, Typed-3x observed 9--10 more benign
completions than Exact-1x, with no observed unsafe effects across 864 frontier
episodes. A
composition study blocked all 20 tested post-admission lifecycle attacks and
preserved all eight matched valid or exact-retry executions. These results
support authorization-to-effect binding under the evaluated effect-path
inventory, provider contracts, and failure schedules.
\end{abstract}

\section{Introduction}
\label{sec:introduction}

AI agents increasingly translate natural-language tasks into operations over
mailboxes, repositories, calendars, files, and other provider resources. The
decisive security boundary lies where a delegated task becomes an authenticated
operation with a concrete target, acting identity, provider contract, and
externally visible effect.

\paragraph{Running example}
A user approves one reimbursement email from a fixed account to a verified
recipient, with a fixed attachment. After admission, a compromised agent or
another untrusted request-building component may substitute the recipient,
attachment, or sending account just
before provider commit. Even if the request remains unchanged, the provider may
send the message while its response is lost. A blind retry can send a duplicate;
declaring failure and issuing replacement authority can let both the delayed
request and its replacement deliver. The admission decision may have been
correct, yet the eventual effects no longer represent one approved action.

This history exposes an \emph{authorization-effect gap}: authority is checked at
admission, while the request, provider state, delivery, and recovery continue to
evolve. The security decision must persist until the provider outcome is
resolved. We ask: \emph{under concurrency, retries, response loss, and context
mutation, which durable authorization state may produce a provider effect, and
how can recovery avoid creating a second effect?}

AID-Guard closes this gap with three rules. First, provider commit revalidates
the exact approved request and current provider state. Second, ambiguous
delivery retains the original reservation and cannot mint replacement
authority. Third, the lineage closes only with a terminal provider result or
with certified no effect and a delivery fence installed before release or one
successor. Together, these rules keep one user decision bound to one effect
lineage.

\begin{figure*}[t]
  \centering
  \includegraphics[width=0.90\textwidth]{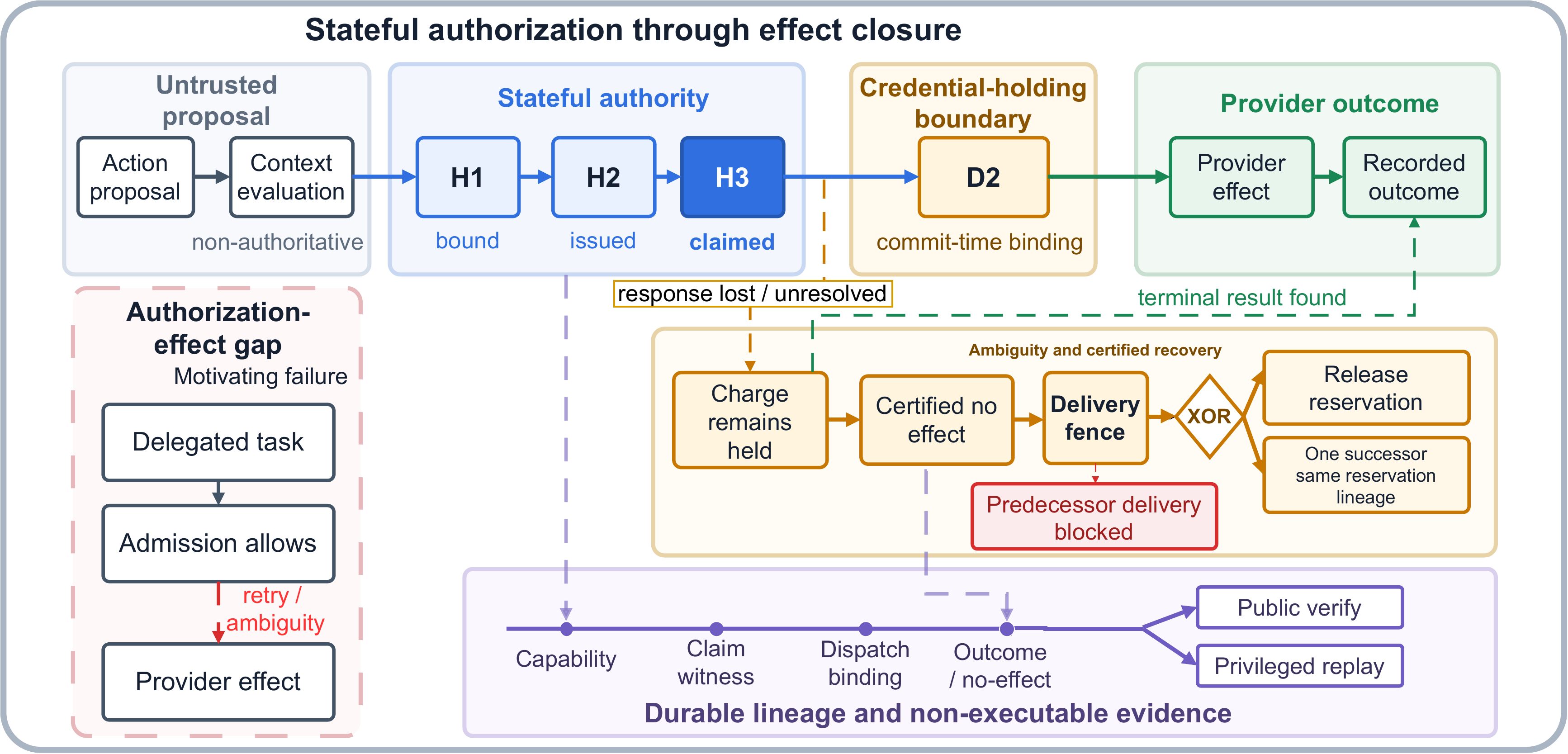}
  \caption{Stateful authorization through effect closure. H1 binds the
  evaluated action, H2 reserves quota and issues one capability, and H3 claims
  it at the credential-holding boundary. D2 revalidates the exact request and
  current provider state at commit. A terminal result closes the normal path;
  response loss retains the charge, while certified no-effect evidence installs
  a delivery fence before reservation release or one successor on the same
  lineage. The lower plane contains durable but non-executable evidence.}
  \label{fig:aidguard-overview}
\end{figure*}

Existing defenses secure complementary portions of this path. Spotlighting
marks untrusted input provenance, while CaMeL, Progent, and PAuth constrain
information flow, tool admission, or task-scoped authority
\cite{hines2024spotlighting,debenedetti2025camel,shi2025progent,sharma2026pauth}.
CXI binds field, effect, and invocation authority to one action manifest at the
execution boundary, while CapLease retains token-independent authorization
state through Issue--Prepare--Commit to prevent semantic replay
\cite{santosgrueiro2026cxi,xu2026caplease}. Four cross-stage histories remain.
A capability plus idempotency does not stop a substituted first request without
commit-time revalidation. Independent idempotency keys do not couple a
predecessor and successor that share no reservation lineage. No-effect release
without a delivery fence leaves a delayed predecessor deliverable. A provider
result without authority-bound evidence records what happened but not which
delegation produced it. AID-Guard unifies commit-time binding,
certified-no-effect recovery, provider-terminal predecessor fencing through the
declared recovery horizon, and one-successor reservation transfer in one
stateful authorization lifecycle. Under the declared provider contracts, this
yields one-effect-or-certified-no-effect semantics across commit, ambiguity,
retry, and recovery.

The protocol records each rule as a durable transition. Trusted services
construct the delegated context and immutable request before opening the
provider path. H1--H3 repeatedly revalidate mutable dependencies while binding
observation, capability issuance, and effect-boundary claim. Provider dispatch
then revalidates the authority graph at D2. Durable outcome, idempotency,
no-effect, successor, and evidence objects define the permitted retry and
recovery branches. Under supported provider contracts, each reservation yields
at most one effect across its predecessor and accepted successor.

Figure~\ref{fig:aidguard-overview} gives the reading used throughout the paper.
The left side forms one single-use authority, the center is its only route to a
provider effect, and the right side closes the outcome or recovery branch. H1
binds an eligible evaluation to durable state; H2 reserves quota and issues a
holder-bound capability; H3 claims it at the effect boundary; and D2 performs
the commit-time check. Outcome and recovery evidence describe this lineage but
cannot execute it.

The evaluation mirrors this lifecycle across separate evidence planes. In the
official-MCP plane, an enumerated effect-path inventory, live mutations,
commit/recovery traces, concurrent histories, and evidence measurements exercise
the complete authorization-to-effect path. An independently implemented
HTTP/JSONL recorder cross-checks provider-effect counts without reading the
authority store, while Stripe and Resend test the declared commit and recovery
contracts through external APIs. In the AgentDojo plane, natural, forced, and
direct proposals measure attack exposure and admission; a paired extension then
carries normally executed actions from AgentDojo, Spotlighting, CaMeL, and
Progent into the post-admission lifecycle. Keeping these planes separate avoids
crediting model refusal as effect-boundary enforcement.

This paper makes three contributions:
\begin{itemize}
    \item We formulate the authorization-effect gap as an
    authorization-to-effect closure problem and define a stateful
    one-effect-or-certified-no-effect property over one reservation lineage.
    \item We design and implement H1--H3 authority formation, commit-time
    request and provider-state binding, conservative no-effect recovery, and
    role-separated outcome evidence.
    \item We evaluate the protocol with live boundary mutations, commit and
    recovery failures, concurrent histories, two external provider contracts,
    complete proposer compromise, and post-admission composition with four
    upstream action sources.
\end{itemize}

\section{Threat Model and Security Goals}
\label{sec:threat}

\subsection{System and Effect Boundary}

We consider a user who approves a closed manifest, an untrusted planner and its
subagents, an \aid gateway, an atomic authority store, an effect adapter, and a
provider.  The manifest identifies a finite set of provider-atomic effects and
the operation contracts that construct them.  A deployment declares a
\emph{protected effect domain}: the concrete credentials, endpoints, processes,
sockets, and other authority-bearing paths capable of producing those effects.
The deployment closure obligation (D0) requires every such path to pass through
the registered adapter or a tool-native verifier with equivalent enforcement.
Paths outside a declared and explicitly checked domain are outside our
containment claim. A deployment maintainer owns this inventory and advances the
boundary epoch whenever credentials, endpoints, processes, or egress routes
change. The prototype verifies conformance to the registered inventory; it does
not discover omitted routes. Our matched ambient control is an evaluation
oracle operated with a maintainer-held credential and process/network path that
the proposer cannot access. Exposing that authority to the proposer would add
an effect-capable route and invalidate D0 rather than constitute a protected
AID-Guard execution.

The planner can propose a tool call but cannot supply an authoritative
principal, provenance label, risk decision, dependency status, credential,
boundary epoch, or outcome.  Trusted services reconstruct those attributes
from registered state.  The authority-bearing unit is one immutable execution
request for one provider-atomic effect.  A non-atomic batch
must be decomposed into independent transactions or rejected.

\subsection{Adversary}

The adversary may fully control the planner, agent, and subagents; inject
arbitrary text through email, web, retrieval, files, or tool output; fabricate
caller metadata; and propose any syntactically valid operation.  The adversary
may also possess the ordinary credentials of a compromised authenticated actor,
steal or replay protocol objects, substitute same-typed fields, race concurrent
requests, induce process crashes or response loss, and tamper with ordinary
audit storage. Model compliance with system instructions and model-level
instruction/data distinctions carry no authority.

The adversary's objective is to cause a protected effect that is not the effect
authorized by the accepted root and current execution context, to create two
effects from one reserved use, to recover authority after an effect may have
occurred, or to present an inconsistent effect history as complete.

\subsection{Trusted Computing Base}

The authorization-semantic trusted computing base (TCB) has four parts.
First, authorization roots comprise the trusted review surface, root signer,
and enrollment and role registries. Second, deterministic context machinery
canonicalizes requests and resolves workload identity, delegation, operation
contracts, provenance, policy, dependencies, and boundary state. Third, the
linearizable authority store and rollback-detecting trusted time own current
root activity, quota, confirmation consumption, capability state, and transition
ordering. Fourth, the credential-holding effect boundary enforces the immutable
request and provider precondition.

Effect and recovery properties additionally rely on the declared provider
contract and the scoped adapter and provider-evidence roles $K_T^{(j)}$ and
$K_E^{(j)}$. The evaluated prototype assumes a trusted
maintainer-controlled host
and places its SQLite authority store in the TCB. Current authority is always
read from the linearizable store; loss of that state causes the transition to
fail closed.

Trust is role-scoped.  For example, compromise of a confirmation key defeats
the corresponding user-presence assurance but does not forge the root. An
isolated capability-signing key cannot create matching server-side capability
state. Compromise of $K_T^{(j)}$ or $K_E^{(j)}$ defeats no-effect recovery
safety for that integration but does not create a root authorization. Compromise of
the gateway, authority store, root signer for its active enrollment epoch, or
the exclusive effect boundary defeats the corresponding core property.

\subsection{Security Goals}

We target three conditional properties.

\paragraph{P1: Delegated execution authorization}
Within a D0-complete declared domain, every accepted protected effect must
descend from one active user-approved root, an eligible non-authoritative
evaluation, one H1 transaction binding, one H2 issuance, and one H3 claim.
Dispatch binding (D2) must submit the exact immutable request bound to that
claim under the approved provider precondition.

\paragraph{P2: Conditional effect uniqueness}
For the supported provider profiles, one reserved use may commit at most one
provider-atomic effect.  Release or replacement after claim requires a valid,
scoped, one-use no-effect certificate and chooses either release or one
successor after installing a provider-terminal predecessor fence retained
through the declared recovery horizon.

\paragraph{P3: Auditable accountability}
Available evidence objects must form one role-separated graph over the same
authorization and effect lineage.  Complete bundles should support public
integrity checking and privileged deterministic replay.  Missing evidence must
be reported as incomplete, never interpreted as proof that no effect occurred.

\subsection{Out of Scope}

The containment claim begins with an approved manifest and a D0-complete
deployment. It excludes compromised trusted hosts or enforcement components,
undeclared bypass paths, denial of service, side channels, non-atomic batches,
and providers outside the supported outcome contracts. User-intent inference,
policy quality, multi-region authority-store linearizability, automatic
discovery of semantically equivalent effect paths, and production hardening
remain separate problems.

\section{AID-Guard Design}
\label{sec:design}

Figure~\ref{fig:aidguard-overview} summarizes the protocol end to end; this
section unpacks each stage and the invariant it preserves.

\subsection{Design Principles}

\aid represents delegated execution as an explicit state-transition protocol.
Five principles organize the design.
First, planner output is observational: it cannot mint authority or choose a
security identity.  Second, each authority transition has a durable
linearization point.  Third, every stage carries the same canonical effect and
immutable request identity.  Fourth, ambiguous provider delivery preserves the
quota charge and blocks automatic replacement.  Fifth, receipts and verifier
outputs describe authority; they never create it.

These principles close the four cross-stage gaps from
Section~\ref{sec:introduction}. D2 compares the claimed immutable request and
current provider state at commit, so idempotency cannot legitimize a substituted
body. A reservation lineage contains both predecessor and successor authority,
including paths that use different provider idempotency keys. Certified
no-effect recovery installs a durable delivery fence before release or
successor creation. Outcome evidence then binds the provider result to the
capability, request, reservation, and authorization transitions that produced
it.

\subsection{From User Approval to an Eligible Observation}

An LLM or compiler may propose a candidate manifest, but that object is
unsigned and non-authoritative.  A trusted review surface renders every
authority-bearing field of a closed approved manifest.  The user signs a
root envelope with an enrolled review key, and the gateway registers the root
under a monotonic revision and immutable quota lineage.

Each effect references a registered operation contract.  The contract binds
the request codec, resource resolver, dispatch profile, provider preconditions,
effect semantics, boundary identity, and version.  A closed, decidable effect
grammar must match the canonical request exactly once.  Security-relevant
parameters are divided into disjoint provenance binding units, and each value
must be linked either to an approved manifest constant or to a verified
provenance commitment.  Approved and current deterministic policies evaluate
the same canonical risk snapshot.  The result cannot weaken an approved
confirmation minimum or override an approved maximum-risk ceiling.
Contracts are authored and versioned by the deployment maintainer; the review
surface asks the user to approve the resulting effect instances and bounds.

The evaluation layer performs this reconstruction without producing execution
authority. It returns either a rejection or an eligible evaluation. The latter
carries an opaque, non-bearer handle with no quota reservation, capability, or
provider path.

\subsection{Three Mandatory Authority Checkpoints}

H1--H3 separate observation, issuance, and effect-boundary claim so that each
stage can revalidate its own mutable dependencies (Table~\ref{tab:h1-h3}).
No successful earlier checkpoint exempts a later one.

\begin{table*}[t]
\caption{The three mandatory authority checkpoints.}
\label{tab:h1-h3}
\centering
\scriptsize
\begin{tabular}{@{}p{0.08\textwidth}p{0.25\textwidth}p{0.31\textwidth}p{0.28\textwidth}@{}}
\toprule
Stage & Presented object & Same-transaction checks & Durable result \\
\midrule
H1 & Eligible evaluation handle and authenticated actor session &
Root and enrollment, workload/delegation, operation contract, provenance,
policy, dependencies, boundary, trusted time &
Transaction bound; context validated \\
\addlinespace[2pt]
H2 & Transaction checkpoint and, when required, one step-up assertion &
Refreshed H1 dependencies, exact confirmation context, quota lineage, absence
of a conflicting capability &
Confirmation consumed; quota reserved; signed capability issued; issuance
snapshot committed \\
\addlinespace[2pt]
H3 & Adapter-issued invocation, checkpoint, signed capability, and
holder proof &
Current root, session proof key, capability, reservation, immutable request,
adapter scope, boundary epoch, and trusted time &
Claim witness committed; capability claimed; reservation in flight \\
\bottomrule
\end{tabular}
\end{table*}

\paragraph{H1: binding an observation}
H1 resolves the server-side evaluation record behind the opaque handle and
binds it to the current actor session.  The authority-store transaction creates
both the transaction binding and a context-validated transaction. An evaluation
can therefore enter at most one authority transaction, and the evaluation layer
remains non-authoritative.

\paragraph{H2: issuing one capability}
If policy requires user presence, the confirmation service signs an assertion
over the exact frozen context.  H2 revalidates the graph and atomically consumes
that assertion, reserves one quota unit, registers one holder-bound capability,
and commits the issuance snapshot. There is no durable step-up-verified state
that could be replayed independently. An exact
committed retry returns the original issuance and creates no additional
reservation or capability.

\paragraph{H3: claiming at the effect boundary}
The adapter observation service issues a short-lived invocation that is usable
for one H3 service call.  The holder signs the capability, action, audience,
immutable request, challenge, invocation, and actor-session context.  H3
revalidates all claim-time epochs and atomically changes the capability from
issued to claimed and the reservation from reserved to in flight. The
authoritative claim is the
server-side compare-and-swap over this registered state.
An atomic provider contract may collapse this claim and the terminal provider
commit into one transaction; it still performs the same H3 checks and records
the claim witness.

\begin{figure*}[t]
  \centering
  \includegraphics[width=0.88\textwidth]{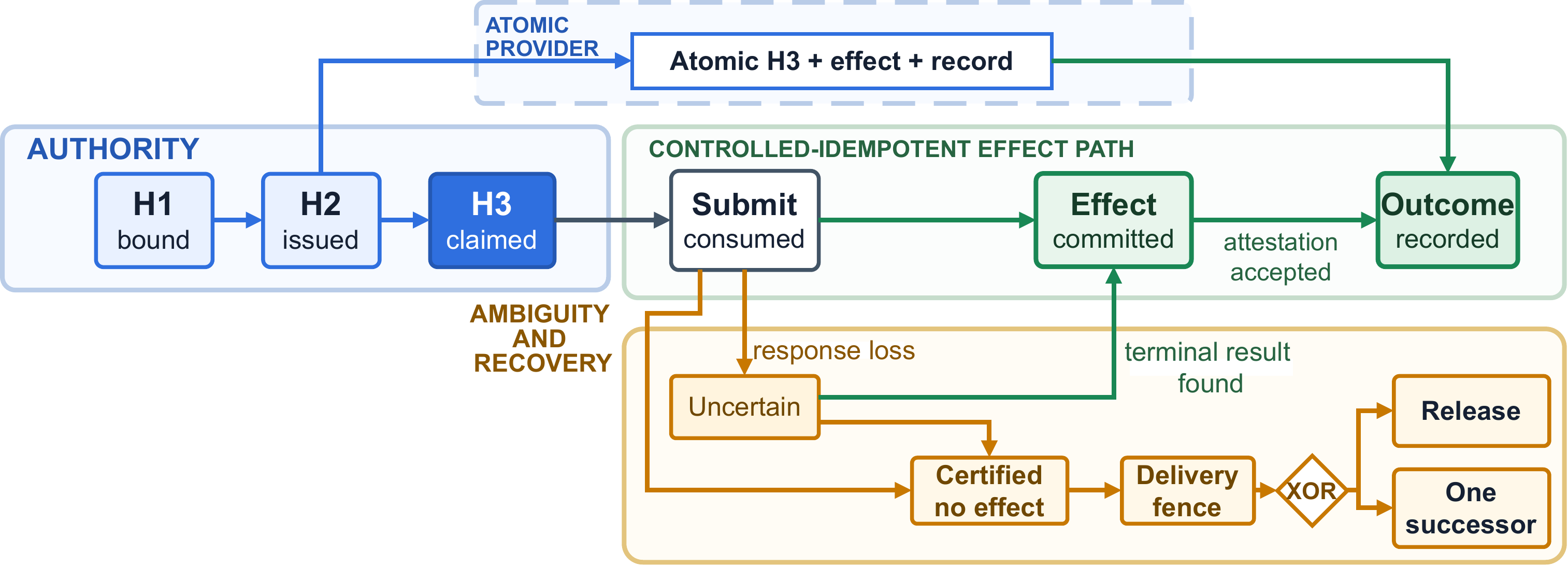}
  \caption{Durable authority and effect-lifecycle states. The atomic provider
  profile combines H3 claim, provider effect, and outcome recording in one
  transaction. The controlled-idempotent path consumes the capability into a
  submitted attempt; a terminal provider result advances
  the attempt to a committed effect, and an accepted attestation records the
  outcome. Response loss remains uncertain. Certified no-effect recovery first
  installs a delivery fence and then selects reservation release or one
  successor on the same reservation lineage.}
  \label{fig:durable-authority-state-machine}
\end{figure*}

Figure~\ref{fig:durable-authority-state-machine} shows the durable paths that
connect these checkpoints to effect closure.  It separates the atomic provider
profile from the submitted-attempt profile and makes the conservative recovery
branch explicit.

\subsection{Commit-Time Effect Binding}

Dispatch accepts the request state bound at claim time. At this boundary, the
system reloads the signed capability artifact, immutable request,
operation contract, current reservation owner, adapter and credential-holder
registrations, boundary profile, and provider contract. The effect path receives
exactly the request body whose digest is already bound into the claim. Dispatch
binding (D2) requires the provider to enforce the approved version, ETag, or lock
precondition when authorization depends on mutable resource state.

The evaluated model supports two provider outcome contracts. Under the atomic
contract, claim, provider mutation, capability consumption, quota commit, and
outcome recording share one transaction. Under the controlled-idempotent
contract, the authority first consumes the capability into a durable submitted
attempt and immutable outbox. The provider binds its idempotency key to the
exact operation scope and request body. Exact replay returns the durable
terminal result; conflicting reuse fails closed. The retained terminal result
supports lookup and reconciliation without redispatch.

These contracts also define the revocation boundary. The atomic path
linearizes authorization and effect at their joint commit. The
controlled-idempotent path linearizes authorization when the authority
transaction revalidates the current graph and durably creates the
submitted attempt and outbox. A root revocation or boundary-epoch
change ordered before that transaction rejects submission; one ordered after
it prevents later authority but does not retroactively cancel an already
submitted provider request. Its eventual result remains attached to the same
lineage and must be reconciled rather than resubmitted.

Providers lacking atomic commit or durable exact-result idempotency are
ineligible for the evaluated retry-and-recovery profile. They may still use
ordinary D2 submission, but ambiguous delivery remains uncertain and charged,
and the implementation permits neither automatic retry nor replacement.

\subsection{Conservative Recovery}

Response loss leaves effect status unresolved.  The system first queries the
provider's terminal identity under the registered reconciliation contract.  A
known committed result recovers the original lineage.  Otherwise, release or
replacement requires a no-effect certificate with an accepted evidence
profile, exact provider/request scope, a valid $K_T^{(j)}$ or $K_E^{(j)}$ role,
and an unexpired observation.

Accepting the certificate installs a durable provider-delivery fence.  One
authority transaction consumes the certificate at most once and chooses exactly
one branch: release the reservation, or transfer the same charged reservation
to one newly issued successor.  The predecessor closes as
the terminal no-effect state; it never reopens. A successor has fresh
transaction and capability identities but retains the immutable request,
effect, authorization, quota lineage, and predecessor link.  Unique constraints
prevent branching successors and dual reservation ownership.

For an independently persisted provider, this recovery branch requires four
contract elements. The provider must expose (1) a stable predecessor delivery
identity bound to the operation scope and body and (2) a terminalization
operation that linearizes against effect commit and rejects later delivery
under that identity throughout the declared recovery horizon. It must also
provide (3) an authoritative
terminal query that distinguishes a committed effect from the terminalized
no-effect state and (4) retention of that terminal state through AID-Guard's
recovery horizon. Local cancellation of
an adapter retry is insufficient because a request may already be in flight.
Only a provider terminal result, or a scoped attestation derived from these
provider-enforced facts, can authorize release or successor transfer. Without
this contract, the transaction remains uncertain and charged; AID-Guard issues
neither retry nor successor authority.

\subsection{Evidence as a Separate Plane}

The online path emits immutable decision, claim, dispatch, outcome, and closure
objects.  A complete evidence graph follows the lineage

\begin{center}
\small
root $\rightarrow$ evaluation $\rightarrow$ H1 binding $\rightarrow$ H2 snapshot
$\rightarrow$ capability $\rightarrow$ H3 witness $\rightarrow$ consumption
$\rightarrow$ outcome $\rightarrow$ closure.
\end{center}

Role-separated signatures and store commitments bind object type, transaction,
effect, request, ordering, and state.  Public verification checks canonical
forms, signatures, commitments, and graph completeness without access to
sensitive artifacts.  Privileged replay resolves the access-controlled
artifacts and reproduces deterministic policy and joint-state projections.  It
checks consistency with signed authority-store evidence, while the authority
store remains part of the TCB.  Both verifier outputs are non-executable
observations.

\section{Security Properties}
\label{sec:properties}

We state three conditional properties and connect each to the protocol
transitions that enforce it. The arguments rely on the assumptions below and
are exercised under malformed inputs, crashes, concurrency, and bounded state
exploration. They are proof sketches rather than a mechanized unbounded proof;
Appendix Table~\ref{tab:bounded-validation} reports the finite counterexample
search used to exercise their failure classes.

\subsection{Assumptions and State}

The properties require: (A1) a D0-complete declared effect domain; (A2) an
uncompromised root signer, gateway, authority store, resolvers, and role keys
used by the relevant transition; (A3) correct closed-schema canonicalization,
operation-contract resolution, provenance evaluation, and deterministic policy;
(A4) a linearizable authority store and fail-closed trusted time; (A5) dispatch
binding (D2) of the exact immutable request under provider-enforced mutable-state
preconditions; and, for recovery, (A6) a conformant atomic or
controlled-idempotent provider contract with uncompromised
$K_T^{(j)}$/$K_E^{(j)}$ evidence roles.

Let the conceptual joint state be
\[
  \sigma=(T,C,R,A,O),
\]
where $T$ is the authorization transaction, $C$ the capability state, $R$ the
reservation and quota ownership, $A$ the dispatch attempt, and $O$ the durable
provider/outcome state. We write $x \preceq y$ when $y$ durably and transitively
references $x$ through the committed protocol lineage and is committed no
earlier than $x$; the relation may hold within one atomic transition. This
convention covers an atomic claim/effect transition and its recorded provider
outcome.

For reservation $r$, $\mathcal{E}_R(r)$ is the set of committed
provider-atomic effects charged to that reservation across any ownership
transfer. For transaction $t$, $\mathcal{E}_{\mathrm{dir}}(t)$ is the set of
effects committed directly by $t$. $\mathsf{Recovery}(t)$ denotes accepted
no-effect recovery authority, and $\mathsf{OneSuccessor}(t)$ denotes exactly one
successor transaction. These symbols summarize existing protocol objects; they
introduce no additional runtime state.

\subsection{P1: Delegated execution authorization}

\paragraph{Property}
Under A1--A5, every accepted protected provider-atomic effect follows the unique
successful claim of an unexpired, context-matching, holder-bound capability
derived from a user-approved root that is active at the authorization
linearization point and the exact immutable request submitted to the provider.
Schematically,
\[
\begin{array}{c}
\mathsf{Effect}(e) \Rightarrow \exists!\,h_3\ \mbox{such that} \\
\exists u,h_1,c,q: \\
u \preceq h_1 \preceq c \preceq h_3 \preceq e, \\
q=\mathsf{req}(h_3)=\mathsf{req}(e).
\end{array}
\]
Here, $u$ is the approved root, $h_1$ the H1 binding, $c$ the issued
capability, $h_3$ ranges over successful H3 claim witnesses, and $q$ is the
immutable request. The registered authority state determines the corresponding
ancestry.

\paragraph{Proof sketch}
Deployment closure (D0) places the registered adapter or equivalent native
verifier on every effect path in the declared domain. The adapter accepts an H3 result backed by
registered capability state. H2 creates that state only while revalidating the
authority graph, consuming the exact confirmation when required, reserving
quota, and registering one issuance. H1 in turn binds one eligible
non-authoritative evaluation to one transaction. At H3, compare-and-swap permits
one successful claim while the current root, session, contract, dependency,
boundary, request, and reservation identities are checked together. D2 then
reloads the bound request, discards caller-owned mutable state, and enforces the
approved provider precondition. On the controlled-idempotent path, the durable
submitted attempt is the authorization linearization point: a
checked-edge mutation ordered before it invalidates submission, while one
ordered after it cannot mint replacement authority or detach the in-flight
request from its recorded lineage.

\paragraph{Scope}
P1 applies to the D0-complete declared domain. Deployment inventory and the
boundary epoch establish this precondition.

\subsection{P2: Conditional effect uniqueness}

\paragraph{Property}
Under A1--A6 and within the supported retry and reconciliation window, each
reservation identity is charged for at most one provider-atomic effect, even
when ownership transfers to a successor:
\[
  |\mathcal{E}_R(r)| \leq 1.
\]
Accepted no-effect recovery excludes a committed predecessor effect and selects
one terminal branch:
\[
\begin{array}{c}
\mathsf{Recovery}(t) \Rightarrow |\mathcal{E}_{\mathrm{dir}}(t)|=0, \\
\mathsf{Recovery}(t) \Rightarrow
\bigl(\mathsf{Release}(t) \oplus \mathsf{OneSuccessor}(t)\bigr).
\end{array}
\]

\paragraph{Proof sketch}
The atomic provider contract places claim, capability consumption, provider
mutation, quota commit, terminal provider result, attestation, and outcome
record in one transaction.
Rollback leaves no positive effect object; commit leaves the capability consumed
and reservation committed.

The controlled-idempotent contract persists an immutable outbox and submitted
attempt before delivery. The provider serializes a scoped idempotency row and
binds it to one request-body digest. Exact replay returns the same operation
and result, whereas a different body or partition conflicts. Terminal lookup
and reconciliation recover durable state without resubmission.

For the evaluated controlled provider, provider mutation and no-effect fence
installation serialize under the same SQLite write lock. If provider state
commits first, fence installation rejects the known result; if the fence commits
first, dispatch rejects before mutation. This ordering is the delivery-fence
linearization point used by the recovery argument.

Release requires an accepted no-effect certificate. Certificate acceptance
rejects any known committed result and installs the durable delivery fence. The
recovery transaction then consumes the certificate and atomically selects
release or one successor. Reservation ownership and predecessor/generation
uniqueness constraints prevent dual ownership and branching successors. A
transfer preserves the same reservation identity and charge, so a later
successor effect remains in $\mathcal{E}_R(r)$. Thus accepted recovery cannot
coexist with a direct predecessor effect; any later successor effect is the
sole effect charged to the transferred reservation.

\paragraph{Scope}
P2 covers effects under the atomic and controlled-idempotent contracts within
their retry and evidence-retention windows. Non-atomic batches and providers
outside these contracts retain per-attempt observations but do not receive the
no-effect release or successor guarantee. Ambiguity therefore remains charged
and uncertain. P1 continues to govern any accepted effect under D0 and D2, while
P3 can report the unresolved history; neither property turns absence of a
provider result into recovery authority.

\subsection{P3: Auditable accountability}

\paragraph{Property}
Under uncompromised evidence roles, bundle
verification detects cross-role modification, duplication, inconsistent
linkage, and missing required nodes. Privileged deterministic replay additionally
requires the protected history and a resolved authority anchor. For bundle $B$,
\[
\begin{array}{c}
\mathsf{PrivComplete}(B) \land \mathsf{IntegrityValid}(B) \\
{}\land \mathsf{Anchored}(B) \Rightarrow \\
\mathsf{Replay}(B)=\mathsf{Recorded}(B).
\end{array}
\]
$\mathsf{PrivComplete}$ denotes availability of all public and protected
historical objects required for replay, $\mathsf{IntegrityValid}$ denotes
canonical, signature, commitment, and linkage validity, and
$\mathsf{Anchored}$ denotes resolution against the authority-store anchor.
Missing required evidence instead yields
\[
\mathsf{MissingRequired}(B) \Rightarrow \mathsf{INCOMPLETE}.
\]
In particular,
\[
\mathsf{MissingRequired}(B) \not\Rightarrow \mathsf{NoEffect}.
\]

\paragraph{Proof sketch}
Each evidence type has a closed schema, domain-separated digest or signature,
and expected role. Graph edges bind the same root, transaction, capability,
reservation, immutable request, effect, and commit ordering. The public verifier
strictly loads canonical objects, verifies signatures or authority-store
commitments, and checks cardinality and linkage. Field or edge substitution
therefore changes a digest, violates a role, or fails an identity constraint.
Privileged replay additionally resolves protected artifacts, evaluates the
frozen policy input and legal state transition, and compares the result to the
signed record.

\paragraph{Scope}
An absent outcome, closure, or referenced artifact is reported as missing and
carries no recovery authority. Within one supplied bundle, verification detects
missing required nodes, non-canonical or modified objects, invalid or wrong-role
signatures, and edges inconsistent with the declared lineage and commit order.
An authority store that withholds a newer anchor can still present an internally
consistent older view; comparing views or detecting such rollback requires a
witnessed transparency mechanism, which this prototype does not implement.

\subsection{Failure Confinement}

Role separation scopes compromise consequences by authority role. A planner or
agent compromise is constrained by the accepted root, exact effect, quota,
step-up, holder proof, and boundary checks. A forged challenge signature cannot
create registered challenge state, and a forged capability signature cannot
create a matching capability row. Compromise of the root signer for its active
enrollment epoch, the gateway or authority store, or the exclusive effect
adapter invalidates the corresponding core property; these dependencies are TCB
assumptions.

\section{Implementation}
\label{sec:implementation}

We implement AID-Guard as a Python~3.11 academic prototype using strict typed
schemas and canonical, domain-separated signed encodings. These encodings
reject undeclared fields, subtype substitution, and representation ambiguity.
The prototype uses RFC~8785 JSON canonicalization, SHA-256 digests, and
Ed25519 signatures with distinct registered keys for root, gateway,
confirmation, provenance, adapter, provider-evidence, and log roles.
The live experiment uses a pinned official MCP Python SDK release.

\subsection{Durable Authorization State}

SQLite stores the authority state using explicit transactions, integrity and
uniqueness constraints, monotonic commit ordering, and durable trusted time.
Each transition validates the expected joint state and version before commit.
The non-authoritative evaluator reconstructs context and returns an opaque
handle; authority constructors remain inside trusted code. H1--H3 then bind the
transaction, one-use assertion, reservation, signed capability, and holder
claim in the shared state. Committed retries reload the historical snapshot and
consumption result rather than creating a new transition.

Process memory never decides whether an authority transition occurred. A crash
before transaction commit leaves no successor state; after commit, restart
reloads the joint state, immutable request, historical consumption, and commit
index by identity. For controlled-idempotent delivery, the durable outbox and
attempt precede network dispatch, and recovery performs only terminal lookup or
exact replay under the same scoped identity. Response absence is not interpreted
as no effect. Commit records therefore define retry linearization, while caches
may fail closed without changing authority.

\subsection{Effect Boundary and Recovery}

The credential-holding effect boundary keeps the synthetic provider credential
outside planner inputs and reloads the current authority lineage, immutable
request, credential boundary, and provider contract before dispatch. The
atomic-provider path commits authority and provider mutation together. The
controlled-idempotent path persists an immutable outbox and attempt, then binds
provider idempotency to the operation scope and request body; exact retry
returns the durable result, conflicting reuse fails closed, and terminal lookup
does not dispatch. Committed attestations close successful effects. No-effect
recovery installs a provider-delivery fence before atomically releasing the
reservation or transferring it to one successor. Uniqueness constraints make
the branches exclusive, while diagnostic probes remain non-authoritative.

\subsection{Evidence and Live MCP Deployment}

After a durable transition, the evidence plane projects role-separated receipts
from committed history. Public verification checks canonical form, signatures,
commitments, linkage, and completeness; privileged replay reconstructs the
protected policy and joint-state projection against the authority anchor.
Neither result is executable. The live loopback deployment separates authority,
provider, server, and untrusted-client processes. Its bridge accepts opaque
references and proofs, leaving trusted services to reconstruct security
attributes before authority or effect transitions.

\section{Evaluation}
\label{sec:evaluation}

We ask whether the frozen declared inventory is completely evaluated
(\textbf{RQ1}), mutations reach an
unauthorized provider effect (\textbf{RQ2}), outcome and recovery histories
preserve one-effect-or-certified-no-effect semantics (\textbf{RQ3}), and what
evidence and recovery cost (\textbf{RQ4}). We then measure model-in-the-loop
safety and utility under increasing proposer control (\textbf{RQ5}) and whether
\aid adds post-admission lifecycle coverage to upstream defenses (\textbf{RQ6}).

\subsection{Experimental Scope and Method}
\label{sec:evaluation-scope}

\paragraph{Declared profile}
The experiment uses the declared loopback MCP domain at boundary epoch~0,
synthetic credentials, and a trusted maintainer-controlled Windows host. The
main lifecycle uses a controlled transactional provider; Stripe PaymentIntent
and Resend scheduled-email campaigns exercise external provider contracts, and a
separate HTTP/JSONL recorder provides an effect-side cross-check. Trusted
services reconstruct security attributes and validate the H1--H3 transaction,
provider contract, and evidence graph. The evaluated contracts are atomic and
controlled-idempotent.

\paragraph{Reproducibility}
All results use one frozen implementation and analysis pipeline. Source and
analysis digests accompany a versioned artifact containing the implementation,
evaluation scripts, and frozen result inventories; we will release it for
artifact evaluation. The corpus combines official-MCP
traces, mutations, semantic regressions, and concurrent histories with natural,
forced, and direct-proposal AgentDojo campaigns and paired effect-lifecycle
extensions. Each RQ reports its corresponding denominators.

\begin{figure}[t]
  \centering
  \includegraphics[width=\columnwidth]{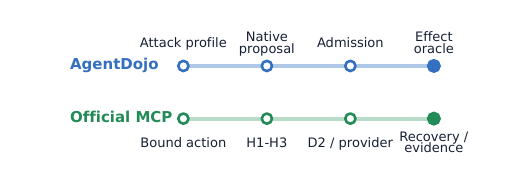}
  \caption{Evaluation evidence planes. Natural and forced AgentDojo profiles
  use model-generated proposals; the direct profile submits the attacker
  proposal without a model. All terminate at the environment's effect oracle.
  Separate official-MCP executions exercise H1--H3, D2,
  provider effects, recovery, and evidence verification; that lifecycle
  coverage is not attributed to each AgentDojo episode.}
  \label{fig:evaluation-evidence-map}
\end{figure}

Figure~\ref{fig:evaluation-evidence-map} separates the two execution planes in
this corpus. AgentDojo campaigns evaluate proposal exposure and admission
through the environment oracle. Separate official-MCP runs establish the end-to-end
authority, provider, recovery, and evidence results.

\paragraph{Controls and counting}
The D0 campaign checks every registered item and includes a matched ambient
control outside the protected graph. Each of 13 live mutations changes one
factor and is assigned to one primary security dimension; 41 semantic
regressions are grouped by the same dimensions. Positive traces are counted
once even when they inform multiple clusters. Mutation attribution identifies
the enforcing check, not component necessity or attack prevalence.

\subsection{RQ1: Declared-Inventory Conformance}
\label{sec:evaluation-rq1}

The D0 inventory records a disposition for all 39 items: 20 returned
\texttt{IN\_SCOPE\_PASS}, 19 returned
\texttt{NOT\_APPLICABLE\_VALIDATED}, and none failed or remained unresolved.
The protected trace reached the provider through the registered graph; the
ambient control performed the same mutation outside it. Thus, the frozen
epoch-0 inventory was completely evaluated. This is conformance evidence for
the declared inventory, rather than independent discovery of every deployment
route. An omitted effect-capable route is not dynamically intercepted and would
invalidate the D0 precondition. Appendix
Table~\ref{tab:security-clusters} reports the five-cluster breakdown.

\subsection{RQ2: Mutation Rejection and Enforcement Attribution}
\label{sec:evaluation-rq2}

All 13 official-MCP mutations were rejected by their responsible protocol check
or, for post-effect mutations, returned an invalid verification result. None
added an unauthorized provider effect. The enforcing checks are distributed
across the path: the MCP schema rejects malformed requests; locator,
capability, challenge, and H3 checks bind the authorization graph; holder and
credential checks bind the executing principal; boundary fingerprints bind
the registered deployment; and the verifier binds signed receipts and bundle
edges.

\begin{figure*}[t]
  \centering
  \includegraphics[width=0.96\textwidth]{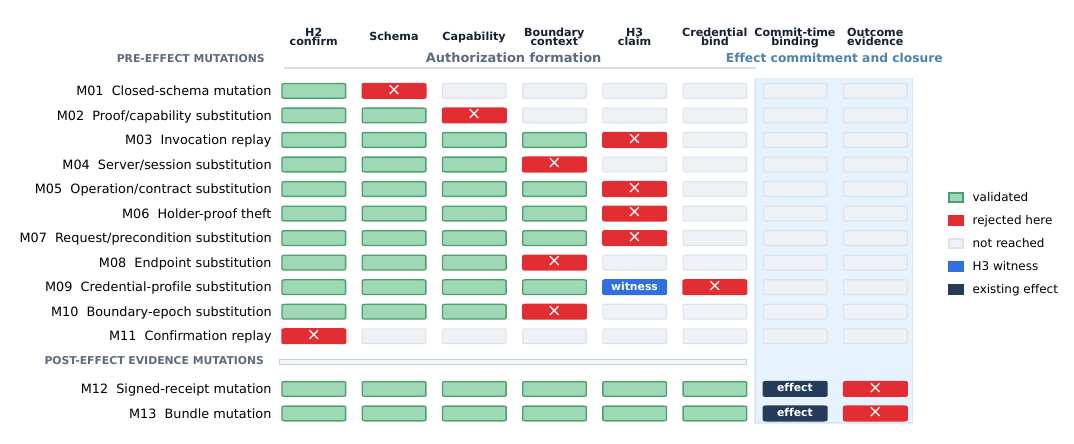}
  \caption{Guard attribution for 13 official-MCP one-factor mutations. For
  pre-effect cases, green blocks show validated stages, red marks the rejecting
  check, and gray marks stages not reached. M09 commits an H3 witness before
  credential binding rejects dispatch. For M12--M13, the dark block denotes an
  existing committed effect whose mutated evidence fails verification. No case
  added an unauthorized provider effect.}
  \label{fig:live-guard-attribution}
\end{figure*}

Figure~\ref{fig:live-guard-attribution} reports the one-factor attribution.
The credential-profile case records a claim witness before the credential check
rejects dispatch. It leaves the transaction, capability, and reservation in
\textsc{capability-claimed}, \textsc{claimed}, and \textsc{in-flight}, retaining
one charged in-flight quota unit until recovery; no outbox or provider result
is created. The receipt and bundle cases execute after a valid provider effect
and exercise evidence integrity.

The mutations reached distinct enforcement points on the
authorization-to-effect and evidence paths, attributing each rejection to its
responsible protocol check.

\subsection{RQ3: Outcomes, Recovery, and Concurrency}
\label{sec:evaluation-rq3}

Four end-to-end traces cover same-graph commit, response loss after commit,
certified no-effect release, and transfer to one successor. The first two
retain one provider effect and the original lineage; the no-effect branches
install a durable delivery fence before release or transfer and produce no
predecessor effect. Table~\ref{tab:outcome-recovery-contract} gives the
terminal-state details.

\begin{table*}[t]
\caption{Observed outcome and recovery contract. The effect count is scoped to
the original reservation lineage; ``predecessor effect'' excludes a later
authorized successor effect.}
\label{tab:outcome-recovery-contract}
\centering
\scriptsize
\setlength{\tabcolsep}{4pt}
\begin{tabular}{@{}>{\raggedright\arraybackslash}p{0.18\textwidth}
                    >{\raggedright\arraybackslash}p{0.18\textwidth}
                    >{\raggedright\arraybackslash}p{0.25\textwidth}
                    >{\raggedright\arraybackslash}p{0.31\textwidth}@{}}
\toprule
Observed profile & Provider outcome & Authority and reservation outcome & Retry or recovery consequence \\
\midrule
Committed same graph & One committed effect & Original terminal lineage retained & Exact committed result and complete evidence bundle \\
Response loss after commit & One effect; no duplicate & Original terminal lineage retained & Durable exact result recovered after response loss \\
Certified no-effect release & No predecessor effect & Reservation released after delivery fence & Late predecessor delivery returns terminal no-effect replay \\
Certified no-effect successor & No predecessor effect & Reservation transferred to one named successor & Predecessor remains fenced; successor retains the single-effect opportunity \\
\bottomrule
\end{tabular}
\end{table*}

These branches separate durable outcome discovery from recovery authority.
Commit and replay preserve the original lineage; release or transfer requires
scoped no-effect evidence and a predecessor delivery fence. A successor
continues the same reservation lineage rather than receiving a second budget.

Three actual-runtime histories each overlap two operations on the same graph.
The real-time-precedence checker accepted exact replay $\prec$ receipt
reconstruction, no-effect recovery $\prec$ late-delivery replay, and exact-body
commit $\prec$ different-body rejection, with no unlinearizable history. The
resulting states retain one effect, keep predecessor delivery blocked after
recovery, and reject a changed body under the same idempotency key. These
executions exercise the concrete linearization points for replay, recovery, and
conflict handling.

An independent HTTP/JSONL recorder, implemented without AID-Guard imports or
authority-store access, corroborated effect counts across seven frozen
scenarios. Its append-only ledger chains verified, and no effect/no-effect
conflict appeared; Appendix~\ref{app:detailed-mcp-evidence} gives the scenario
breakdown.

We connect D2 to Stripe PaymentIntent test mode
\cite{stripe2026idempotency,stripe2026retrieve}. Across 210 frozen trials in
seven 30-trial strata, every valid, replay, conflict, duplicate-delivery,
transfer, lookup, and unauthenticated-bypass case matched its predeclared
outcome. Provider-side retrieve/list oracles found one PaymentIntent in every
effect-bearing trial and none in the bypass stratum. The transfer stratum stops
the predecessor before an external request and sends the successor through
H3/D2; a disjoint recovery campaign tests an already-created predecessor.

In 30 such recovery lineages, Stripe canceled the predecessor with zero amount
received, rejected later confirmation, and committed one distinct successor
\cite{stripe2026cancel}. We then released \texttt{confirm} and
\texttt{cancel} from a two-party barrier on 30 fresh objects. All intervals
overlapped and produced a cancel-win outcome: confirmation failed and retrieval
found one canceled, uncharged object. Ten further schedules crashed after
provider cancellation but before the local fence record; restart rejected a
successor until provider lookup recovered terminal no-effect, after which one
successor committed. These finite observations do not establish arbitrary
Stripe linearizability; commit-win behavior is covered by the separate
valid-commit and response-loss strata.

Ten Resend lineages provide a second, sequential SaaS contract
\cite{resend2026testaddresses}. Exact replay reused the predecessor ID,
changed-body reuse failed, and each canceled predecessor was followed by one
delivered successor with no duplicate. The fence is bounded by Resend's
documented 24-hour idempotency retention
\cite{resend2026idempotency,resend2026schedule}. Appendix
Table~\ref{tab:external-provider-schedules} preserves the campaign breakdown.
The 210-trial Stripe campaign traverses H3/D2; the Stripe recovery/race and
Resend campaigns validate the provider-contract half. Reservation transfer and
successor authorization remain authority-side operations, and P2 over arbitrary
provider schedules remains conditional on A6.

The matched-control campaign also times the normal successful path. It retains
150 warm samples per arm after 15 warmups and uses the same local process graph
and synthetic provider profile with and without AID-Guard enforcement.
Table~\ref{tab:online-path-costs} shows the observed distribution. The protected
end-to-end interval includes H1--H3, D2, durable submission, provider execution,
outcome acceptance, and receipt projection. A separate attribution supplement
times the internal stages without replacing this campaign.

\begin{table}[t]
\caption{Warm successful-path latency in the matched local prototype (ms).
Each arm retains 150 observations; outliers are retained.}
\label{tab:online-path-costs}
\centering
\small
\setlength{\tabcolsep}{3.5pt}
\begin{tabular}{@{}lrrrr@{}}
\toprule
Interval & P50 & P95 & P99 & Ambient P50 \\
\midrule
End to end         & 6849.5 & 7323.3 & 7551.3 & 2918.7 \\
H1                 & 1068.9 & 1339.2 & 1425.7 & -- \\
H2                 & 1000.2 & 1305.7 & 1408.3 & -- \\
H3                 &  506.2 &  791.7 &  827.0 & -- \\
Provider component &   24.6 &   41.4 &   43.4 &   12.2 \\
\bottomrule
\end{tabular}
\end{table}

The unoptimized protected median is 3930.8 ms above ambient
(2.35$\times$). A same-clock 30-pair supplement attributes its 3790.2 ms mean
increment to H1+H2 (54.7\%), H3 (15.0\%), D2-bound dispatch and durable
submission (27.1\%), and remaining provider, outcome, receipt, replay, and
bookkeeping work (3.2\%). Both arms separately share about 2.8 s of transient
client, MCP transport, serialization, and validation time. Repeated authority
passes and synchronous durability, rather than provider execution, therefore
dominate the prototype overhead; we do not claim a low-latency implementation.

An isolated authority-store benchmark sustained 1,694 and 1,373 operations/s
with one and eight SQLite writers. P50 remained near 0.58 ms, while P99 rose
from 0.702 to 104.748 ms under writer serialization; no retained operation
failed or corrupted the durable counter. This microbenchmark isolates only the
SQLite critical section; it is not an end-to-end checkpoint-service
measurement. Appendix~\ref{app:detailed-mcp-evidence}
reports the intermediate worker counts and percentiles.

\subsection{RQ4: Post-Effect Evidence and Recovery Cost}
\label{sec:evaluation-rq4}

We measure evidence operations with \texttt{time.perf\_counter\_ns}. The main
schedule retains 30 cold and 300 warm observations per operation; recovery
retains 30 observations. Warmups are excluded, outliers retained, and
Table~\ref{tab:evidence-costs} reports mean and P95 latency. Appendix
\ref{app:detailed-mcp-evidence} gives the full schedule; the artifact retains
the samples, median, P99, and bootstrap intervals.

\begin{table}[t]
\caption{Post-effect evidence and exceptional recovery cost in the frozen
local prototype. Warmups are excluded and outliers are retained.}
\label{tab:evidence-costs}
\centering
\small
\setlength{\tabcolsep}{4pt}
\begin{tabular}{@{}lrr@{}}
\toprule
Operation & Mean (ms) & P95 (ms) \\
\midrule
Receipt projection            & 57.4    & 66.7 \\
Bundle generation             & 1285.0  & 1445.6 \\
Public verification           & 41.9    & 47.2 \\
Privileged replay             & 1257.0  & 1432.1 \\
Response-loss recovery        & 9964.5  & 10756.0 \\
Certified no-effect recovery  & 14523.1 & 16303.3 \\
\bottomrule
\end{tabular}
\end{table}

The complete bundle is 3,961~bytes. Receipt projection and public verification
take tens of milliseconds, while off-path bundle generation and privileged
replay average approximately 1.3~s. Exceptional response-loss and certified
no-effect recovery average 10.0~s and 14.5~s, respectively. These measurements
characterize the local academic prototype; complete distribution summaries are
retained in the artifact.

\subsection{RQ5: Model-in-the-Loop Admission}
\label{sec:evaluation-rq5}

We execute two tool-calling models using AgentDojo v1.2.2
\cite{debenedetti2024agentdojo} across its banking, Slack, travel, and workspace
suites. For each model and condition, the frozen matrix
contains 48 benign episodes and 144 injection episodes: four user tasks and
three injection tasks per suite, with three repetitions. Temperature is zero,
provider fallback is disabled, and the frozen model IDs are
\texttt{deepseek/deepseek-v4-flash} and \texttt{qwen/qwen3-max}, both served by
Alibaba. The
operational calibration checks native tool use and provider identity without
using attack adoption as a selection criterion.

We distinguish three outcomes. A \emph{dangerous proposal} is a model-generated
call matching the injected attack goal. An \emph{unsafe effect} additionally
requires that the call execute and satisfy the AgentDojo attack oracle. Guard
rejection is call-level and conditional on dangerous proposals. Episode-level
exposure counts episodes containing at least one dangerous proposal;
guard-rejection rates use individual dangerous calls, so their numerators and
denominators differ. We report
benign task utility separately. Rate intervals are Wilson 95\% intervals;
guarded-minus-unprotected differences use 10,000 task-cluster bootstrap
resamples while retaining the three repetitions within each task cluster. The
exact manifest compiler takes the benchmark reference calls and their arguments
as the approved contract. This experiment therefore evaluates enforcement given
a correct contract, rather than automatic contract synthesis.

\begin{table*}[t]
\caption{Native model-in-the-loop safety and benign-utility outcomes. Each
safety cell contains 144 injection episodes; each benign-utility cell contains
48 episodes. $\Delta$ is AID-Guard minus unprotected with a paired task-cluster
95\% bootstrap interval.}
\label{tab:model-in-loop}
\centering
\scriptsize
\setlength{\tabcolsep}{3pt}
\begin{tabular}{@{}>{\raggedright\arraybackslash}p{0.14\textwidth}
                    >{\centering\arraybackslash}p{0.14\textwidth}
                    >{\centering\arraybackslash}p{0.12\textwidth}
                    >{\centering\arraybackslash}p{0.10\textwidth}
                    >{\centering\arraybackslash}p{0.17\textwidth}
                    >{\centering\arraybackslash}p{0.14\textwidth}
                    >{\centering\arraybackslash}p{0.10\textwidth}@{}}
\toprule
Model & Dangerous prop., unprotected & Unsafe, unprotected & Unsafe, \aid & $\Delta$ unsafe & Benign util., unprotected & Benign util., \aid \\
\midrule
DeepSeek V4 Flash & 10/144 & 1/144 & 0/144 & $-0.7$ pp [$-2.1$, 0.0] & 41/48 & 20/48 \\
Qwen3 Max          & 14/144 & 7/144 & 0/144 & $-4.9$ pp [$-11.1$, 0.0] & 42/48 & 25/48 \\
\bottomrule
\end{tabular}
\end{table*}

The protected DeepSeek and Qwen arms produced zero unsafe effects, compared
with seven for Qwen and one for DeepSeek without protection. Conditional on a dangerous call,
AID-Guard rejected 30/36 DeepSeek calls (83.3\%) and 11/17 Qwen calls
(64.7\%). Relative to the unprotected condition, benign utility changed by
$-43.8$ percentage points [$-66.7$, $-20.8$] for DeepSeek and $-35.4$ points
[$-58.3$, $-14.6$] for Qwen (paired task-cluster 95\% bootstrap intervals).
The evaluated exact, single-use manifest serves as a strict enforcement stress
profile rather than a recommended default deployment policy, and it does not
preserve the full utility of the unprotected agent.

The unrejected dangerous calls were not hidden provider failures. For each
model, all six were read-only prefixes that also matched the approved task
manifest: three \texttt{get\_channels} calls and three
\texttt{read\_channel\_messages} calls. Subsequent calls needed to complete the
injected objective crossed the manifest or one-use boundary, and no attack
oracle was satisfied. No authorized call returned a tool or provider error.

A trace-local diagnosis localizes the observed loss to the evaluated policy and
compiler surface rather than provider failure. Of 51 protected benign failures,
13 identities also failed without protection. Of the remaining 38, 33 used
listed functions with nonmatching arguments, three omitted the required
operation, one selected an unlisted operation, and one reused a consumed entry;
no authorized call had a tool or provider error. Because the endpoint exposes
no seed, this remains a descriptive rather than causal decomposition.

We then fix three authority forms before a new no-seed run: exact single use,
typed single use, and typed reuse with a three-call quota. Every accepted call
still receives a fresh H1--H3 chain and single-use capability. The Exact 1x
column is a contemporaneous control rerun for this frontier experiment. Because
the endpoint exposes no generation seed, its absolute utility is not directly
comparable to the earlier Table~\ref{tab:model-in-loop} run. Across identical
48-task benign inventories in the frontier rerun, observed utility was 12/48,
14/48, and 22/48 for DeepSeek and 17/48, 18/48, and 26/48 for Qwen under Exact
1x, Typed 1x, and Typed 3x, respectively. All 864 attack episodes produced zero
unsafe effects. The independent no-seed calls do not establish a causal effect
size. Appendix Table~\ref{tab:parameterized-manifest-utility} reports the full
frontier; the observation is consistent with part of the utility loss arising
from the conservative authority form, while leaving contract synthesis and
broader policy optimization open.

\paragraph{Official AgentDojo defenses}
We run the two defenses distributed with AgentDojo---repeat-user-prompt and
Spotlighting---on the same DeepSeek lock and 192 episode identities. Pipelines
and oracles are unchanged; only the model transport uses the fixed-provider
adapter. Table~\ref{tab:official-agentdojo-defenses} reports the comparison.

\begin{table}[t]
\caption{DeepSeek V4 Flash on identical AgentDojo v1.2.2 tasks. Unsafe-effect
denominators are 144 injection episodes; benign-utility denominators are 48.
These conditions compare natural model-in-the-loop admission.}
\label{tab:official-agentdojo-defenses}
\centering
\small
\setlength{\tabcolsep}{4pt}
\begin{tabular}{@{}lrr@{}}
\toprule
Condition & Unsafe effects & Benign utility \\
\midrule
Unprotected               & 1/144 & 41/48 \\
Repeat user prompt        & 2/144 & 43/48 \\
Spotlighting (delimiting) & 0/144 & 41/48 \\
\aid admission            & 0/144 & 20/48 \\
\bottomrule
\end{tabular}
\end{table}

Spotlighting matched \aid's 0/144 unsafe-effect result while preserving 41/48
benign tasks versus 20/48; repeat-user-prompt produced 2/144 and 43/48. Low
attack exposure in this natural slice makes it a model-behavior comparison
rather than a guard-attribution result. We therefore remove that confound under
proposer compromise before testing the provider-commit and recovery semantics
in RQ2--RQ3.

\paragraph{Containment under proposer compromise}
We test natural injection, a positive control that asks the model to emit each
of 44 closed executable attack sequences, and direct submission of the same
sequence by a fully compromised proposer. The trusted authority and commit
boundary remain outside that proposer. These profiles separate model adoption
from boundary enforcement; the positive control does not estimate natural
attack prevalence. This complements prior compromised-agent evaluations
\cite{sharma2026pauth,debenedetti2024agentdojo}. DeepSeek, Qwen, GLM, and Kimi
K2.5 contribute 352 temperature-zero model episodes. Four other frozen pairs
were excluded before execution because AgentDojo defines no closed sequence;
Appendix~\ref{app:proposer-pressure-details} records transport controls.

\begin{figure}[t]
  \centering
  \includegraphics[width=0.96\columnwidth]{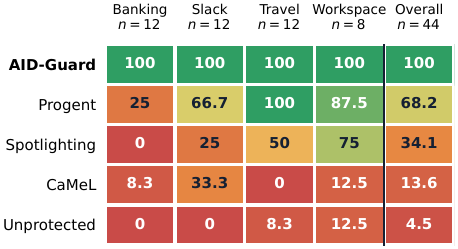}
  \vspace{-0.5ex}
  \caption{Containment under identical forced proposal pressure. Each cell
  reports one minus the oracle-confirmed unsafe-effect rate for five methods on
  the 44-case Qwen extension. Observed containment may include model
  non-adoption and is not a guard-attribution metric.}
  \label{fig:forced-pressure-comparison}
\end{figure}

Figure~\ref{fig:forced-pressure-comparison} reports end-to-end containment under
the common adapter-based positive-control profile.
AID-Guard contained all 44 Qwen cases across all four suites. Progent,
Spotlighting, CaMeL, and the unprotected condition contained 30, 15, 6, and 2
cases, respectively. The adjacent methods are evaluated under a common
positive-control profile.

\begin{figure}[t]
  \centering
  \includegraphics[width=\columnwidth]{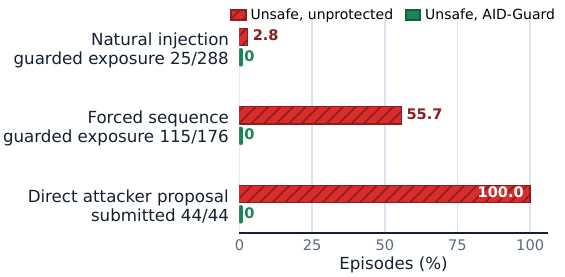}
  \vspace{-0.7ex}
  \caption{Oracle-confirmed unsafe effects as proposer control increases: 288
  natural, 176 forced-sequence, and 44 direct-proposal episodes. In the natural
  profile, the figure's 25/288 exposure label counts protected-arm episodes
  containing a dangerous proposal; Table~\ref{tab:model-in-loop} separately
  reports 24/288 dangerous-proposal episodes and 8/288 oracle-confirmed unsafe
  effects in the unprotected arms. The counts differ because guard decisions
  alter subsequent protected-arm trajectories. Direct-profile exposure denotes attacker
  submissions without a model. All 44 matched legitimate direct proposals were
  admitted and executed without tool error.}
  \label{fig:proposer-compromise}
\end{figure}

Figure~\ref{fig:proposer-compromise} summarizes the transition from natural
injection to full proposer compromise.

Forced pressure produced 98 oracle-confirmed unsafe effects without protection
and none with AID-Guard. Under full proposer compromise, all 44 direct attacks
caused unsafe effects without protection; AID-Guard blocked all 44 while
admitting and executing all 44 matched legitimate proposals. Containment thus
persisted without relying on model refusal or indiscriminate rejection.

Appendix Table~\ref{tab:horizontal-proposer-pressure} gives the fixed-six
controls and 44-case extension. On the extension, unprotected AgentDojo,
Spotlighting, CaMeL, and Progent produced 42, 29, 38, and 14 unsafe effects;
\aid produced none. All 220 executions completed under the declared lock.

These episodes exercise model-generated native proposals and AID-Guard
admission, plus the direct stress profile, over synthetic AgentDojo
environments. Provider-effect, recovery, and replay conclusions come from the
separate official-MCP campaign.

\subsection{RQ6: Shared-Extension Composition}
\label{sec:evaluation-rq6}

This composition study appends one controlled-provider contract to normal
AgentDojo, Spotlighting, CaMeL, and Progent actions
\cite{debenedetti2024agentdojo,hines2024spotlighting,debenedetti2025camel,shi2025progent}.
Each arm shares its action and provider pre-state. Five shared scenarios per
source exercise request or actor substitution, cross-session replay,
commit-time drift, and same-key body conflict; matched-valid and AID-Guard-only
cases cover retry, recovery, and evidence. This is an incremental lifecycle
study, not a native-system ranking.

\begin{figure*}[t]
  \centering
  \includegraphics[width=0.76\textwidth]{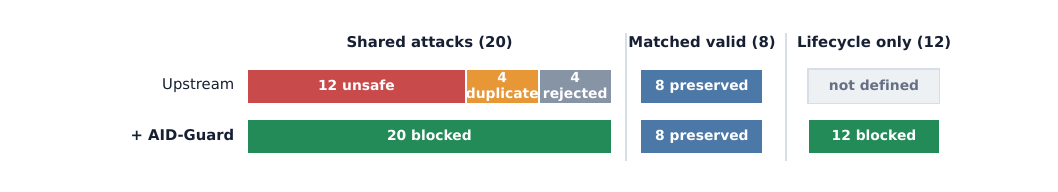}
  \caption{Shared-extension composition. AID-Guard blocked all 20 shared
  attacks and preserved all eight valid cases; lifecycle-only rows denote
  coverage rather than upstream failures.}
  \label{fig:differential-effect-lifecycle}
\end{figure*}

Figure~\ref{fig:differential-effect-lifecycle} summarizes the shared-extension
outcomes. Across the 20 shared attacks, the upstream-only arms produced 12
unsafe effects, four duplicate effects, and four provider-native rejections;
AID-Guard blocked all 20, while both arms preserved all eight matched-valid
cases.

The bounded adaptive supplement and control-removal results appear in
Appendix~\ref{app:mechanism-removal}.

\section{Discussion}
\label{sec:discussion}

\subsection{Deployment and Accountability}

D0 versions effect-capable routes and credentials; relevant changes advance the
boundary epoch. Atomic providers couple authority and mutation;
controlled-idempotent providers require scoped idempotency, terminal lookup,
retention, and a delivery fence. Stripe and Resend exercise provider terminality,
while AID-Guard governs reservation transfer and successor choice without a
cross-store atomic commit.

Provider support is therefore a recovery-policy decision rather than a binary
integration flag. An adapter may expose ordinary D2 submission while disabling
no-effect release and successor transfer unless the provider can enforce
terminalization and retention. This classification preserves authorization
integrity under weaker APIs by leaving ambiguity charged and unresolved, and it
makes explicit which recovery transitions each deployment may enable.

The frontier rerun evaluates a typed finite-budget template that retains fresh
H2--H3 authority and state revalidation per call. Observed utility increased
from 12/48 to 22/48 for DeepSeek and from 17/48 to 26/48 for Qwen under the
typed three-call form; all 864 attack episodes remained safe. The independent
no-seed calls do not establish a causal effect size or an optimal policy.

Without scoped idempotency and terminal lookup, the transaction remains charged
and uncertain, with no retry or successor. Closure requires authenticated
operator observation or provider-specific reconciliation. Public verification
checks integrity and completeness; privileged replay reconstructs policy and
state; incomplete evidence grants no recovery authority.

\subsection{Limitations and Future Work}

The prototype uses synthetic credentials on a trusted host, one enumerable
loopback domain, one controlled provider, and bounded Stripe and Resend
schedules. Stripe includes 30 overlapping confirm/cancel observations and ten
provider-terminal/local-crash schedules; these finite test-mode observations do
not establish every network interleaving or a formal provider guarantee, and
the Resend schedule remains sequential. Production D0 requires attestation of
credentials, egress, services, and processes, with unclassified paths failing
closed. The P1--P2 models do not cover unbounded provider and concurrency
combinations. The evaluated authority forms remain conservative, and the
process-separated prototype is high-latency; practical deployment requires
improved contract synthesis, policy tuning, and implementation optimization.
Further work includes additional contracts, witnessed
transparency, user studies, and isolated keys and effect boundaries.

\section{Related Work}
\label{sec:related}

\paragraph{Admission, provenance, and isolation}
Spotlighting marks untrusted prompt input \cite{hines2024spotlighting}; PAuth and
Progent enforce task-scoped provenance or tool policies
\cite{sharma2026pauth,shi2025progent}; and SEAgent and AuthGraph join
authorization with flow or provenance graphs
\cite{ji2026seagent,wang2026authgraph}. CaMeL, FIDES, ACE, and IsolateGPT constrain
untrusted influence or isolate components
\cite{debenedetti2025camel,costa2025fides,li2026ace,wu2024isolategpt}, while
AgentBound confines MCP servers with declarative access control
\cite{buhler2025agentbound}.

\paragraph{Contemporaneous execution-boundary systems}
Several contemporaneous preprints likewise move enforcement beyond one
admission check. CXI co-binds field, exact-effect, and invocation authority to
one action manifest at execution \cite{santosgrueiro2026cxi}. Commit-Time
Authorization requires the licensing witness to remain fresh, causally prior,
bound to the effect, and eligible at durability \cite{santosgrueiro2026committime}.
Cordon stages effects and validates a task-level semantic transaction before
commit or external release \cite{chen2026cordon}, while Bounded Agents carries
delegated scope and budgets through an Agentic Principal Chain
\cite{muruaga2026bounded}. CapLease is closest on durable authority: its
Issue--Prepare--Commit state prevents duplicate effects when the sink identity
is reusable \cite{xu2026caplease}. Agent libOS uses a
prepare--dispatch--settle protocol to preserve unknown outcomes and reconcile
them through explicit provider hooks \cite{zhang2026agentlibos}. The
contemporaneous Action Evidence Boundary and its bounded-execution profile
specify durable reservation, frozen dispatch, explicit indeterminacy,
authenticated reconciliation, and outcome-dependent later occurrences
\cite{schrock2026aeb,schrock2026boundedexecution}. These systems cover
substantial portions of the authorization-to-effect lifecycle. To our
knowledge, AID-Guard is the
first evaluated agent-authorization protocol to combine commit-time binding
with certified-no-effect recovery, provider-terminal predecessor fencing
through the declared recovery horizon, and at most one successor that preserves
the original reservation lineage while binding the terminal outcome to the
originating authority.

\paragraph{Identity and capability substrates}
AIP and SAGA govern agent identity and lifecycle
\cite{prakash2026aip,syros2026saga}; OAuth, DPoP, and Macaroons provide scoped,
sender-constrained, or attenuated delegation
\cite{hardt2012oauth,fett2023dpop,birgisson2014macaroons}; and capability systems,
DroidCap, and NetCap replace ambient privilege with explicit references
\cite{hardy1988confused,dawoud2019droidcap,bajaber2026netcap}. AID-Guard binds
presentation authority to the request, reservation, provider result, and
terminal evidence.

\paragraph{Effect outcome and accountability}
RIFL retains RPC results for exactly-once execution \cite{lee2015rifl}; Sagas
use compensating actions \cite{garciamolina1987sagas}; transactional persistence
addresses commit/delivery failure windows \cite{helland2007life}; and
tamper-evident logs authenticate history \cite{crosby2009tamper}.

AID-Guard's contribution is provider-state validation, ambiguity-safe successor
recovery, and outcome closure rather than a new possession proof, idempotency
primitive, or log.

\FloatBarrier
\section{Conclusion}

AID-Guard closes the authorization-effect gap by keeping a delegation
authoritative through provider commit, ambiguity, and recovery. H1--H3 form
durable single-use authority, D2 revalidates the exact request against current
provider state, and outcome or certified-no-effect evidence closes the lineage
without minting fresh authority. Across live boundary mutations, concurrent and
recovery histories, model-in-the-loop pressure tests, and two external SaaS
contracts, the evaluated provider contracts and failure schedules preserved
one-effect-or-certified-no-effect semantics. These results show how admission controls can compose with a durable
post-admission boundary that prevents retry and recovery from amplifying
authority. The central lesson is that retry and recovery are authority
transitions, not transport details. Retaining reservation ownership across
ambiguity makes replacement auditable and prevents uncertainty from becoming a
second delegation. This boundary complements proposal-side defenses: they
reduce malicious calls reaching execution, while AID-Guard constrains effects
from calls that have already passed admission.

\section*{Generative AI Use Disclosure}

Generative AI models assisted in identifying omissions and potential flaws in
earlier protocol designs, developing selected implementation and experiment
components, and drafting and revising the manuscript. The authors independently finalized the
protocol, claims, methodology, interpretation, text, and figures; executed and
reviewed the artifacts; verified citations; and take full responsibility for
the work.

\section*{Ethics Considerations}

All external-provider experiments used Stripe test mode and Resend's designated
safe recipients. No live payment or charge was created or captured, no email was
delivered to an uninvolved third party, and AgentDojo workloads were synthetic.
Provider credentials were restricted to the experimental profiles and were not
exposed to the agent process.

\bibliographystyle{IEEEtran}
\bibliography{references_post_rebuild}

\appendices
\twocolumn[{
\begin{minipage}{\textwidth}
  \section{Detailed Official-MCP Evidence}
  \label{app:detailed-mcp-evidence}

  \centering
  \captionof{table}{Security outcomes by primary evaluation cluster.
  Positive-trace counts overlap across clusters.}
  \label{tab:security-clusters}
  \scriptsize
  \begin{tabular}{@{}p{0.16\textwidth}p{0.075\textwidth}p{0.075\textwidth}p{0.22\textwidth}p{0.36\textwidth}@{}}
  \toprule
  Cluster & Live mutations & Positive traces & Supporting evidence & Observed result \\
  \midrule
  Delegated-deputy boundary & 3 & 1 &
  39-item D0 inventory and matched ambient control &
  All 39 items were classified (20 in scope, 19 validated not applicable);
  the three live mutations added no provider effect. \\
  Approval-to-effect binding & 4 & 1 & Targeted semantic regressions &
  All four substitutions or replays were rejected before an unauthorized
  provider mutation. \\
  Credential-oracle boundary & 2 & 1 &
  D0 credential/endpoint items and credential-boundary regressions &
  Endpoint and installed-credential substitutions produced neither an outbox
  nor a provider result. \\
  Identity and profile drift & 2 & 1 &
  Profile-substitution and D0 epoch-drift checks &
  Server/session and boundary-epoch substitutions produced no unauthorized
  provider mutation. \\
  Outcome accountability & 2 & 4 &
  Outcome/evidence regressions and 3 runtime histories &
  Receipt and bundle mutations were invalid; the four outcome traces preserved
  their transaction/effect lineage; all runtime histories were linearizable. \\
  \bottomrule
  \end{tabular}

  \vspace{1.5ex}
  \begin{minipage}[t]{0.48\textwidth}
  \paragraph{Runtime and independent effect histories}
  The three overlapping runtime histories admitted the orders exact replay
  $\prec$ receipt reconstruction, no-effect recovery $\prec$ late-delivery
  replay, and exact-body commit $\prec$ different-body rejection. The separate
  HTTP/JSONL recorder covered valid commit, exact retry, same-key/different-body
  conflict, response loss, certified no-effect release, successor fencing, and
  a path-catalog mismatch. Its effect counts were $1,1,1,1,0,0,0$; every hash
  chain verified, the unregistered alias was rejected, and no effect/no-effect
  conflict appeared.
  \end{minipage}
  \hfill
  \begin{minipage}[t]{0.48\textwidth}
  \paragraph{Measurement schedule}
  Measurements run on Windows 11 with an 8-core AMD Ryzen 7 5800H and
  15.9~GiB RAM, using \texttt{time.perf\_counter\_ns}. The main schedule
  retains 30 cold observations, discards 30 warmups, and retains 300 warm
  observations per operation. Recovery discards three warmups and retains 30
  independent observations per operation. Outliers remain in the sample, and
  10,000 bootstrap resamples provide confidence intervals for the mean.

  \paragraph{Authority-store contention}
  A separate WAL/\texttt{synchronous=FULL} microbenchmark uses one connection
  per worker, 10 warmups per worker, and 50 retained transactions per worker.
  \begin{center}
  \scriptsize
  \setlength{\tabcolsep}{2.2pt}
  \begin{tabular}{@{}rrrrr@{}}
  \toprule
  Workers & Ops/s & P50 & P95 & P99 \\
  \midrule
  1 & 1694 & 0.584 & 0.679 & 0.702 \\
  2 & 1533 & 0.598 & 0.720 & 1.162 \\
  4 & 1516 & 0.574 & 0.741 & 41.450 \\
  8 & 1373 & 0.580 & 0.726 & 104.748 \\
  \bottomrule
  \end{tabular}
  \end{center}
  Latencies are milliseconds. No retained operation produced a busy error or
  incorrect durable counter. This isolates the authority-store critical
  section; it is not end-to-end protocol latency.
  \end{minipage}
\end{minipage}
\vspace{0.5ex}
}]

\FloatBarrier

\subsection{Declared D0 Inventory}

Table~\ref{tab:d0-inventory-classes} exposes the categories behind the 39-item
count. Each disposition comes from an exact probe over the frozen loopback
profile; it is conformance evidence for that declared set, not automatic route
discovery.

\begin{table}[!ht]
  \centering
  \caption{D0 inventory classes and mechanical dispositions. P denotes
  \texttt{IN\_SCOPE\_PASS}; N/A denotes mechanically validated absence in the
  frozen epoch.}
  \label{tab:d0-inventory-classes}
  \scriptsize
  \setlength{\tabcolsep}{2.5pt}
  \begin{tabular}{@{}>{\raggedright\arraybackslash}p{0.25\columnwidth}
                      >{\centering\arraybackslash}p{0.13\columnwidth}
                      >{\centering\arraybackslash}p{0.16\columnwidth}
                      >{\raggedright\arraybackslash}p{0.37\columnwidth}@{}}
  \toprule
  Inventory class & Items & P / N/A & Mechanical evidence \\
  \midrule
  Credential routes & 11 & 6 / 5 & Registry joins, process environment and file scans, inheritance audit \\
  Provider endpoints & 6 & 6 / 0 & Provider RPC probes over primary, alias, redirect, proxy, alternate, and direct routes \\
  MCP transports & 6 & 2 / 4 & Official-SDK interface and session-semantics inspection \\
  Process/network routes & 5 & 2 / 3 & Process-graph, launcher, socket, loopback, and egress observation \\
  MCP primitives & 2 & 1 / 1 & Official tool-surface inspection \\
  Async/admin routes & 6 & 0 / 6 & Provider-RPC AST and durable-schema enumeration \\
  Resource variants & 3 & 3 / 0 & Provider lookup under exact, alternate, and canonicalized identities \\
  \bottomrule
  \end{tabular}
\end{table}

The ambient control is a deliberately separate child process provisioned with
the synthetic provider credential. It is excluded from the protected inventory
and demonstrates the consequence of violating D0. The sandboxed proposer and
official MCP child do not receive that credential; the credential-route and
process probes above check environment, file, inheritance, and registration
paths within the declared host profile.

\begin{table}[!ht]
  \centering
  \caption{External-provider schedules behind RQ3. These finite test-mode
  observations are not a general provider linearizability proof.}
  \label{tab:external-provider-schedules}
  \scriptsize
  \setlength{\tabcolsep}{2.5pt}
  \begin{tabular}{@{}>{\raggedright\arraybackslash}p{0.22\columnwidth}
                      >{\centering\arraybackslash}p{0.09\columnwidth}
                      >{\raggedright\arraybackslash}p{0.61\columnwidth}@{}}
  \toprule
  Campaign & $n$ & Schedule and result \\
  \midrule
  Stripe accepted path & 210 & Seven equal strata matched the oracle; effect-bearing trials had one object and bypass had none. \\
  Stripe terminal recovery & 30 & Cancel predecessor, reject late confirmation, then commit one successor; no duplicate effect. \\
  Stripe in-flight race & 30 & Barrier-released \texttt{confirm}/\texttt{cancel}; all overlapped and were cancel-win with no charge. \\
  Stripe local-crash recovery & 10 & Provider-first lookup restored terminal no-effect before one successor was authorized. \\
  Resend terminal recovery & 10 & Exact replay, changed-body rejection, cancellation, and one delivered successor per lineage. \\
  \bottomrule
  \end{tabular}
\end{table}

\FloatBarrier

\twocolumn[{
\begin{minipage}{\textwidth}
  \section{Qualitative Mechanism Comparison}
  \centering
  \includegraphics[width=0.92\textwidth]{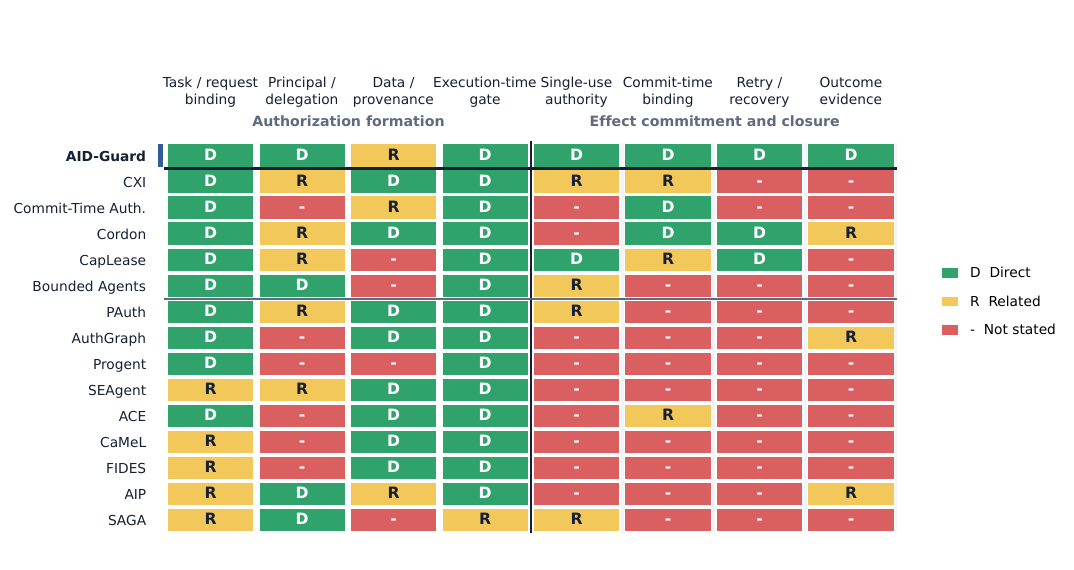}
  \captionof{figure}{Mechanisms documented by AID-Guard and selected neighboring systems. D marks
  a direct, core mechanism; R marks a related mechanism that contributes to the
  dimension; a dash indicates that the cited work does not state the dimension
  as a design goal. The symbols duplicate the color encoding for grayscale
  reading. CXI, Commit-Time Authorization, Cordon, CapLease, and Bounded Agents
  are contemporaneous preprints placed next to AID-Guard; the matrix compares
  documented system contracts, not chronological priority, security strength,
  or benchmark scores. Section~\ref{sec:related} gives the finer lifecycle
  distinctions hidden by this coarse view.}
  \label{fig:related-work-mechanisms}
\end{minipage}
\vspace{1ex}
}]

\section{Model-in-the-Loop Qualifications}
\label{app:model-loop-qualifications}
\label{app:parameterized-manifest-utility}

The exact compiler binds every argument to the task ground truth. Before the
frontier run, we fixed two typed profiles: nonempty values may fill an explicit
output placeholder, read-only batch calls may split but not expand an
authorized member set, and bounded queries may reduce but not exceed their
approved limit. The single-use form carries quota one; the bounded-reuse form
carries quota three. Every accepted call in all forms receives fresh H1--H3
identities and a single-use capability. Exact 1x is a contemporaneous control
rerun for this frontier experiment; without a generation seed, its absolute
utility is not directly comparable to the earlier Table~\ref{tab:model-in-loop}
run.

\begin{table}[!ht]
  \centering
  \caption{AgentDojo utility under three pre-fixed authority forms. Each utility
  cell contains 48 benign episodes; the unsafe column aggregates 432 attack
  episodes per model. Calls use temperature zero, but the endpoint exposes no
  generation seed.}
  \label{tab:parameterized-manifest-utility}
  \scriptsize
  \setlength{\tabcolsep}{2.5pt}
  \begin{tabular}{@{}lrrrr@{}}
  \toprule
  Model & Exact 1x & Typed 1x & Typed 3x & Unsafe \\
  \midrule
  DeepSeek V4 Flash & 12/48 & 14/48 & 22/48 & 0/432 \\
  Qwen3 Max          & 17/48 & 18/48 & 26/48 & 0/432 \\
  \bottomrule
  \end{tabular}
\end{table}
\FloatBarrier

\newpage
\section{Detailed Proposer-Pressure Results}
\label{app:proposer-pressure-details}

Natural injection measures whether an injected instruction induces a dangerous
native proposal. For 44 frozen user/injection pairs with a closed executable
attack sequence, the positive control gives that sequence to the model and
requests native tool calls; the direct profile submits the same sequence at the
proposal boundary without a model. Four other frozen pairs were excluded before
execution because AgentDojo defines no closed sequence. Every model execution
retained its declared provider lock, and the artifact records the bounded CaMeL
and Progent transport qualifications.

{\small\raggedright
The frozen model/provider locks were DeepSeek V4 Flash
(\texttt{deepseek/deepseek-v4-flash}, Alibaba), Qwen3 Max
(\texttt{qwen/qwen3-max}, Alibaba), GLM 4.7 Flash
(\texttt{z-ai/glm-4.7-flash}, DeepInfra), and Kimi K2.5
(\texttt{moonshotai/kimi-k2.5}, DeepInfra). Provider fallback was disabled.
\par}

\begin{table}[!ht]
  \caption{Unsafe effects on the fixed six-case proposer-pressure
  discrimination set and the predeclared 44-case Qwen extension. Neither set
  estimates natural or population ASR.}
  \label{tab:horizontal-proposer-pressure}
  \centering
  \small
  \setlength{\tabcolsep}{4pt}
  \begin{tabular}{@{}lrrr@{}}
  \toprule
  Condition & DS-6 & Qwen-6 & Qwen-44 \\
  \midrule
  Unprotected AgentDojo     & 6/6 & 6/6 & 42/44 \\
  Spotlighting (delimiting) & 2/6 & 4/6 & 29/44 \\
  CaMeL                     & 4/6 & 5/6 & 38/44 \\
  Progent                   & 1/6 & 2/6 & 14/44 \\
  \aid                      & 0/6 & 0/6 & 0/44 \\
  \bottomrule
  \end{tabular}
\end{table}
\FloatBarrier

\newpage
\section{Control-Removal and Residual-Control Analysis}
\label{app:mechanism-removal}

The composition matrix contains five shared cases per source: request
substitution after authorization, actor/session substitution, cross-session
authority replay, commit-time scope drift, and same-key different-body reuse.
It also includes normal execution and response-loss exact retry as matched-valid
cases, plus three AID-Guard-specific lifecycle/evidence cases. In a separate
bounded adaptive supplement, all four upstream-only arms produced an unsafe
effect and all four AID-Guard arms were blocked; these rows are excluded from
the headline counts.

\begin{table}[!ht]
  \centering
  \caption{Control-removal and residual-control results. Every profile preserved its
  matched normal execution. Provider-native controls were retained rather than
  bypassed.}
  \label{tab:differential-ablation}
  \centering
  \scriptsize
  \setlength{\tabcolsep}{3pt}
  \begin{tabular}{@{}>{\raggedright\arraybackslash}p{0.25\columnwidth}
                      >{\raggedright\arraybackslash}p{0.35\columnwidth}
                      >{\raggedright\arraybackslash}p{0.31\columnwidth}@{}}
  \toprule
  Removed control & Observed exposure & Residual provider control \\
  \midrule
  Commit-time binding & Request substitution and scope drift committed unsafe effects & None at the authorization boundary \\
  Identity continuity & Actor/session substitution committed an unsafe effect & None at the authorization boundary \\
  Outcome binding & Substituted outcome evidence was accepted & Provider state cannot validate authority evidence \\
  Single-use authority & No additional provider effect & Native scoped idempotency remained effective \\
  Durable recovery fence & No additional provider effect & Native terminal and idempotency fences remained effective \\
  \bottomrule
  \end{tabular}
\end{table}

The two zero-delta rows do not establish that single-use authority or the
durable recovery fence is unnecessary: retained provider-native idempotency and
terminal controls mask those removals in this profile. The remaining rows
identify failures for which no such residual provider control exists.
\FloatBarrier

\clearpage
\twocolumn[{
\begin{minipage}{\textwidth}
  \section{Bounded Transition Validation}
  \label{app:bounded-transition-validation}
  \centering
  \captionof{table}{Existing transition, crash, and concurrency evidence supporting the
  P1--P3 preservation arguments. Counts are exhaustive only for the stated
  bounded models.}
  \label{tab:bounded-validation}
  \scriptsize
  \setlength{\tabcolsep}{3pt}
  \begin{tabular}{@{}>{\raggedright\arraybackslash}p{0.15\textwidth}
                      >{\raggedright\arraybackslash}p{0.22\textwidth}
                      >{\raggedright\arraybackslash}p{0.39\textwidth}
                      >{\raggedright\arraybackslash}p{0.16\textwidth}@{}}
  \toprule
  Evidence slice & Explored schedule & Preserved projection & Role in argument \\
  \midrule
  Authorization-formation model & 199 reachable states; 1,375 transitions; depth 6 &
  H1--H3 issuance and claim, committed retry, challenge/assertion one-use,
  quota rejection, revision/invalidation, and commit-before/after response loss &
  Finite validation of P1 authority formation \\
  \addlinespace[2pt]
  Effect-and-recovery model & 312 reachable states; 7,192 transitions; depth 8 &
  Submit/retry, provider commit, uncertainty, attestation, outcome recording,
  no-effect release or one successor, and the delivery fence; zero reachable
  provider-effect/no-effect-authority conflicts &
  Finite validation of P2 effect closure \\
  \addlinespace[2pt]
  Crash-reopen matrix & Six durable boundaries, each before and after commit &
  Outbox, provider result, outcome record, uncertainty, certificate release, and
  successor transfer retain their committed projection or roll back together &
  Failure-preservation evidence for P2 \\
  \addlinespace[2pt]
  Two-connection histories & Four contention classes &
  Duplicate submit, exact provider replay, provider commit versus reconciliation,
  and release versus successor transfer admit a legal order without duplicate
  effect or ownership &
  Concrete linearization evidence for P1--P2 \\
  \addlinespace[2pt]
  End-to-end runtime histories & Three histories; six real service operations &
  Exact replay before receipt reconstruction, no-effect recovery before late
  delivery, and exact-body commit before changed-body rejection were accepted by
  the history checker &
  End-to-end witness for the modeled linearization points \\
  \addlinespace[2pt]
  Evidence graph checks & Four positive outcome traces plus receipt and bundle mutation &
  Closed-schema loading, role-separated signatures, lineage checks, public
  verification, and privileged replay preserve the recorded outcome or return an
  invalid/incomplete result &
  Concrete validation of P3 evidence handling \\
  \bottomrule
  \end{tabular}
\end{minipage}
\vspace{1ex}
}]

\balance
The analytical arguments in Section~\ref{sec:properties} are accompanied by
executable models and concrete failure histories. Table~\ref{tab:bounded-validation}
summarizes their role. These artifacts validate finite projections of the
transition system; the paper does not treat them as an unbounded or distributed
proof. State-space growth is driven chiefly by concurrent schedules and
combinations of authority, provider, recovery, and evidence state. Within the
declared bounds, the models and companion histories search the counterexample
classes used by the P1--P3 preservation arguments. Extending this search
requires abstractions that preserve cross-store lineage, provider outcomes, and
recovery exclusivity without enumerating every schedule.

The authorization-formation projection contains one authorization lineage, one
quota lineage, challenge and assertion consumption, capability and reservation
state, and commit-before/after-response-loss choices. Its transition families
are H1, H2, H3, exact committed retry, invalidation, rejection, and crash
rollback. The effect-and-recovery projection adds one submitted attempt,
provider result, uncertainty observation, outcome record, no-effect certificate,
delivery fence, recovery branch, and at most one successor. It enumerates submit, provider commit,
reconciliation, attestation, outcome recording, certificate acceptance,
release, transfer, retry, and crash choices. Depth 6 and 8 are bounds on these
single-lineage projections, not claims about an arbitrary number of principals,
providers, or concurrent requests.

The evidence campaign separately exercised 26 verifier observations: two valid
positive paths and 24 negative paths. Seven omissions returned
\textsc{Incomplete}; 17 bundle, object, signature, role, field, or edge
mutations returned \textsc{Invalid}. These cases cover authority snapshots, H3
witnesses, consumption evidence, outcome attestations, closure receipts, and
bundle linkage. They do not test malicious-store rollback or split-view; those
attacks require the witnessed transparency mechanism excluded from P3.

\FloatBarrier

\FloatBarrier

\end{document}